\documentclass[a4paper,11pt]{article}
\usepackage{pos}
\usepackage[acronym]{glossaries}
\makeglossaries
\usepackage{graphicx}
\usepackage{wrapfig}
\usepackage{multicol}
\usepackage{slashed}
\usepackage{physics}
\usepackage{subcaption}
\usepackage{wrapfig}
\usepackage{microtype}
\usepackage{upgreek}
\usepackage[colorinlistoftodos,prependcaption,textsize=small]{todonotes}

\newacronym{hic}{HIC}{Heavy-Ion Collision}
\newacronym{ipr}{IPR}{Inverse Participation Ratio}
\newacronym{lqcd}{LQCD}{Lattice QCD}

\title{QCD Anderson transition at zero and non-zero external magnetic fields}

\author*[a]{R. Kehr}
\author[b]{A. Dean M. Valois}
\author[a,c]{L. von Smekal}

\affiliation[a]{Institut für Theoretische Physik, Justus-Liebig-Universität,
Heinrich-Buff-Ring 16, 35392 Gießen, Germany}

\affiliation[b]{CAFPE and Departamento de Física Teórica y del Cosmos, Universidad de Granada, E-18071
Granada, Spain}

\affiliation[c]{Helmholtz Forschungsakademie Hessen für FAIR (HFHF), GSI Helmholtzzentrum für
Schwerionenforschung, Campus Gießen}

\emailAdd{Robin.Kehr@physik.uni-giessen.de}

\abstract{The QCD Anderson transition is believed to be connected to both deconfinement and chiral crossovers. These crossovers are substantially affected when external magnetic fields ($B$) are present, most prominently, e.g., via magnetic catalysis and inverse magnetic catalysis. In this work, we use lattice QCD to investigate the Anderson transition in two different setups: (1) at $B=0$ by studying the low-lying eigenmodes of the overlap operator using gauge configurations with $2+1+1$ quark flavors of twisted-mass Wilson fermions. We estimate the mobility edge below which eigenmodes are localized via the inflection point of the so-called relative volume. Previous work has shown that, contrary to expectations, this estimate does not vanish at the temperature of the chiral phase transition. A possible scenario for this apparent contradiction was discussed, and in this work, we present an alternative observable for measuring localization that supports this scenario. And (2) by studying the localization properties of the staggered Dirac operator at $B\neq0$ on configurations with $2+1$ dynamical staggered fermions and 2 stout-smearing steps. Our preliminary results on two lattice spacings ($24^3\times 6$ and $24^3\times 8$) indicate a non-monotonic behavior of the mobility edge with the magnetic field across different temperatures, which hints at a reduction in the Anderson transition temperature in the presence of an external magnetic field.}

\FullConference{The 42nd International Symposium on Lattice Field Theory (LATTICE2025)\\
 2 November - 8 November 2025\\
Tata Institute for Fundamental Research, Mumbai, India\\}

\begin{document}
\maketitle

\section{Introduction}\label{sec:introduction}

The QCD Anderson transition~\cite{Giordano:2021qav} is the QCD analog of the Anderson transition in condensed matter physics~\cite{Anderson:1958vr, Evers:2007zsx} and is believed to be connected to the deconfinement \cite{Kovacs:2017uiz, Holicki:2018sms} and chiral \cite{Giordano:2014pfa, Holicki:2018sms, Giordano:2022ghy, Kehr:2023wrs, Kehr:2025nth} transitions. In contrast to condensed matter physics -- where the Anderson transition emerges in the localization properties of the eigenmodes of the Hamiltonian -- in QCD, the localization manifests itself in the eigenmodes of the Dirac operator. Below a certain temperature -- which is expected to be roughly around the chiral transition temperature $T_\mathrm{c}$ -- all modes are delocalized and well-described by random matrix theory.
Above this temperature, the lower-lying modes are localized, following a Poisson distribution, while the higher ones are delocalized. The spectral position where this transition occurs is called the mobility edge. A commonly used measure of localization is the relative volume occupied by an eigenmode in spacetime, which reads
\begin{equation}
	r(\lambda) = \frac{{\rm IPR}_2^{-1}(\lambda)}{N_\mathrm{s}^3 N_\mathrm{t}}\in [0,1]\,,\hspace{0.3cm}\text{with}\hspace{0.3cm}{\rm IPR}_q(\lambda)  = \sum_{n \in \Lambda} ( v_\lambda(n)^\dagger  v_\lambda(n))^q\,,  \label{eq:eigenmode_relvol}
\end{equation}
where $q\in\mathbb{Z}$ is the order of the \gls{ipr} of an eigenmode $v_\lambda$, and $n$ denotes a given lattice site. Here, $\Lambda$ denotes the set of all lattice sites. The mobility edge can then be estimated by the inflection point $\lambda_\mathrm{c}$ of the relative volume, which can be determined by fitting a polynomial of a given order to the data. 

Another subject of phenomenological interest is the influence of external magnetic fields on the properties of the strong interactions. Strong fields are known to be produced in the early stages of peripheral \glspl{hic}, inside neutron stars, and are also conjectured to have appeared during the electroweak phase transition in the early Universe (see Ref.~\cite{Adhikari:2024bfa} for a broad review). The impact of magnetic fields on QCD has also been extensively studied using lattice methods, including their impact on the approximate QCD order parameters and on the Equation of State (see Ref.~\cite{Endrodi:2024cqn} for a review). Some of the most prominent effects of magnetic fields on the strong interactions are the enhancement of the chiral condensate at $T \ll T_\mathrm{c}$ and at $T \gg T_\mathrm{c}$ (magnetic catalysis), the suppression of the condensate for $T \approx T_\mathrm{c}$ (inverse magnetic catalysis), and the reduction of the $T_\mathrm{c}$~\cite{Bali:2011qj}. Due to the conjectured connection between chiral and Anderson transitions, it is insightful to understand how these effects manifest in terms of the localization properties of the Dirac operator at non-zero magnetic fields. However, to the best of our knowledge, little is known about the influence of magnetic fields on the Anderson transition. Hence, in this work, we give the first steps towards filling this gap using \gls{lqcd}.

This proceedings article consists of two main parts: in the first one (Sec.~\ref{sec:overlap_tmft}), we give an update on previous studies by some of us in a mixed setup with overlap valence quarks on a twisted mass sea at vanishing magnetic field~\cite{Kehr:2023wrs, Kehr:2025nth}. In the second part (Sec.~\ref{sec:influence_magnetic_fields}), we study the influence of magnetic fields on the Anderson transition using staggered fermions both in the valence and in the sea sectors. Finally, in Sec.~\ref{sec:conclusions}, we summarize our findings and discuss the outlook of the project.

\section{Update for the setup with overlap valence quarks on a twisted-mass sea}\label{sec:overlap_tmft}

In this section, we briefly summarize the setup of our previous studies in order to provide an update with more recent results. For more details and previous results, we refer the reader to Refs.~\cite{Kehr:2023wrs, Kehr:2025nth}. In this setup, we compute the low-lying eigenmodes of the overlap operator~\cite{Neuberger:1997fp}
\begin{equation}
	D_{\mathrm{ov}} = \frac{\rho}{a} \, (1 + \mathrm{sgn}\,K) \,,
\end{equation}
where the kernel operator $K=a D_\mathrm{W}-\rho$ is given by the Wilson operator $D_\mathrm{W}$ with the shift set to $\rho=1.4$ in order to optimize locality according to Ref.~\cite{Cichy:2012vg}. The gauge configurations, which were generated with $N_\mathrm{f} = 2+1+1$ flavors of twisted-mass Wilson fermions at maximal twist and the Iwasaki gauge action, were taken from the \emph{twisted mass at finite temperature} (tmfT) collaboration \cite{Burger:2018fvb}. The parameteres were taken from several references, namely, the lattice spacing $a = 0.0619(18)\,\mathrm{fm}$ from Ref.~\cite{EuropeanTwistedMass:2014osg}, the pion mass $m_\uppi = 225(7)\,\mathrm{MeV}$ from Ref.~\cite{ExtendedTwistedMass:2019omo}, and the pseudocritical temperature $T_\mathrm{pc}=171(6)\,\mathrm{MeV}$ from Ref.~\cite{Kotov:2021rah}. The number of lattice sites in each space direction $N_\mathrm{s}=48$ results in a lattice extent of $L = 2.97(9)\,\mathrm{fm}$ and the temperature is varied by the number of lattice sites in the temporal direction $N_\mathrm{t}\in\{4,6,\dots,18,20,24\}$. After computing the eigenvalues, they are additionally stereographically projected from the Ginsparg-Wilson circle onto the imaginary axis, as a means of estimating the effect of continuum extrapolations.

Employing the relative volume~\eqref{eq:eigenmode_relvol}, we found a non-vanishing mobility edge for all evaluated temperatures, even at the lowest temperature of $T=133(4)\,\mathrm{MeV}$, corresponding to $N_\mathrm{t}=24$. This coincides with the independently determined temperature of the chiral phase transition in the chiral limit of $T_\mathrm{chiral}=132^{+3}_{-6}\,\mathrm{MeV}$ \cite{HotQCD:2019xnw}, where we expect all eigenmodes to be delocalized~\cite{Giordano:2022ghy}. A possible scenario for this apparent contradiction was already discussed in previous publications~\cite{Kehr:2023wrs, Kehr:2025nth}. In these references, we argued that measuring $r(\lambda)$ at a fixed volume may be insufficient, since an eigenmode with a small relative volume could still scale according to some nonzero effective dimension when varying the volume, meaning that it would not be genuinely localized. Evidence for this scenario and the existence of a second infrared mobility edge is given by studies in quenched and full QCD \cite{Alexandru:2021pap, Alexandru:2023xho, Meng:2023nxf}.

\begin{figure}[!ht]
    \centering
    \begin{subfigure}{0.49\textwidth}
        \includegraphics[width=\linewidth]{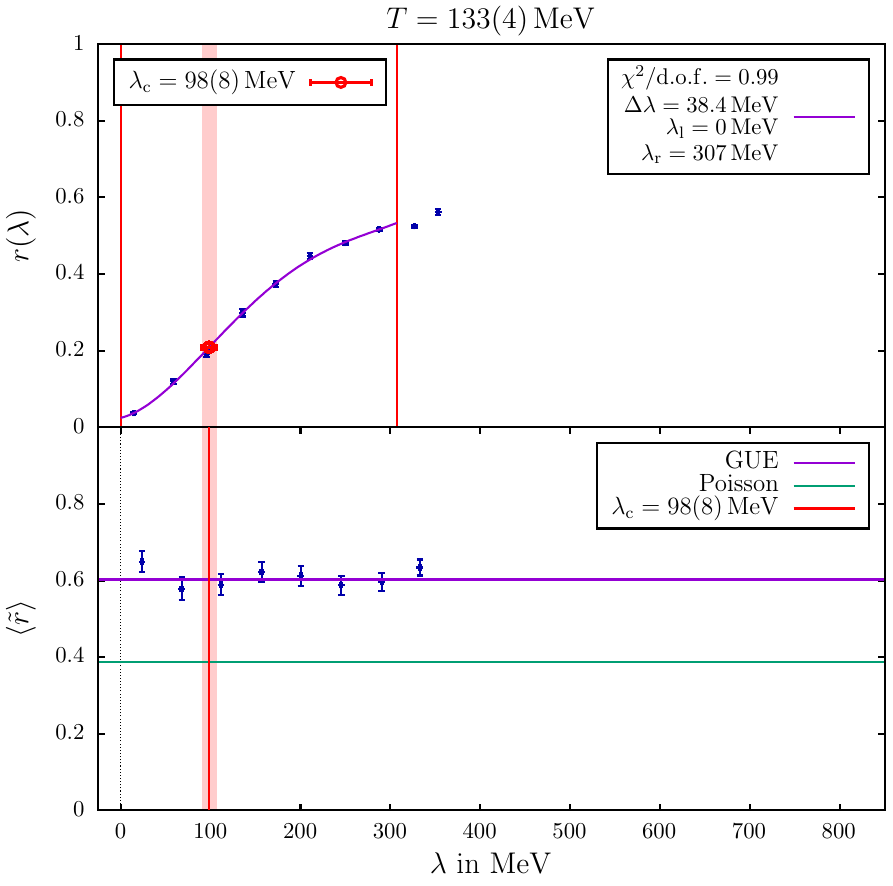}
    \end{subfigure}
    \begin{subfigure}{0.49\textwidth}
        \includegraphics[width=\linewidth]{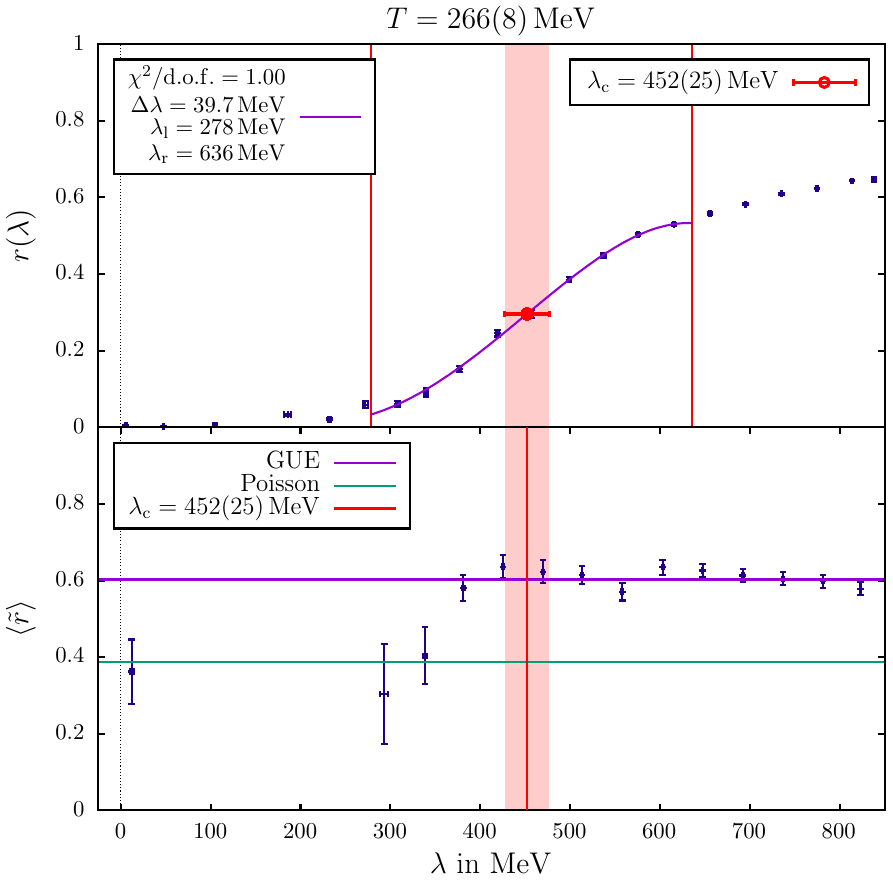}
    \end{subfigure}
    \caption{Expectation value of $\tilde r$ (bottom), defined in Eq.~\eqref{eq:r_tilde}, compared to the relative volume (top) as functions of $\lambda$ for $T=133(4)\,\mathrm{MeV}$ ($N_\mathrm{t}=24$) and $T=266(8)\,\mathrm{MeV}$ ($N_\mathrm{t}=12$). The data points for $\langle \tilde{r} \rangle$ result from averaging over small bins of size $\Delta\lambda \approx 45\,\mathrm{MeV}$. The vertical red-shaded bar highlights the error range of the mobility edge estimate $\lambda_\mathrm{c}$, which is extracted from the inflection point of $r(\lambda)$, including statistical and systematic (estimated by a second fit) error. The fit window $[\lambda_{\mathrm l}, \lambda_{\mathrm r}]$ is indicated by vertical red lines in the plot of $r(\lambda)$, while the vertical line in the plot of $\langle \tilde{r} \rangle$ displays the position of $\lambda_\mathrm{c}$.}
    \label{fig:r_tilde}
\end{figure}

However, an extensive analysis of the volume dependence of $r(\lambda)$ is currently not feasible in our setup due to the high costs of the overlap operator. Therefore, it is worth investigating other observables that could capture the localization behavior. One possibility is the quantity~\cite{Pandey:2024goi, Oganesyan:2007wpd}
\begin{equation}
	\tilde r_n = \min\left( \frac{s_{n+1}}{s_n}, \frac{s_n}{s_{n+1}} \right) \in [0,1] \,,
    \label{eq:r_tilde}
\end{equation}
where $s_n = \lambda_{n+1} - \lambda_{n}$ denotes the so-called level spacing of two consecutive eigenvalues. This is based on the observation that the unfolded level spacings of delocalized eigenmodes are distributed according to the Gaussian unitary ensemble, while the ones of the localized eigenmodes follow Poisson statistics \cite{Kovacs:2012zq}. It follows that~\cite{Atas:2013gvn}
\begin{equation}
	\langle \tilde{r} \rangle \approx 
    \begin{cases}
		0.603  \quad \textrm{for Gaussian unitary ensemble (GUE),} \\
		0.386 \quad \text{for Poisson distribution}\,.
		\end{cases}
\end{equation}
Therefore, this quantity is a candidate for a measure of localization. In Fig.~\ref{fig:r_tilde}, we show results for $\langle \tilde{r} \rangle$ together with $r(\lambda)$ at two temperatures.
According to $\langle \tilde{r} \rangle$, there are no localized modes at the chiral phase transition temperature (in the chiral limit), which can be seen in the plot on the left-hand side showing the lowest temperature coinciding with $T_\mathrm{chiral}$. In contrast, the plot on the right-hand side at a higher temperature shows localized modes for the lower-lying eigenvalues, as expected\footnote{It is worth noting that the estimate based on the mobility edge may have weaknesses, since it seems to overestimate the transition point when compared to $\langle \tilde{r} \rangle$. Nevertheless, the lattice artifacts on each of these observables need to be further investigated.}. However, the data points have large error bars, especially for the localized modes, due to their low occurrence. This also applies to the intermediate temperatures, where the onset of localization is expected. Therefore, improving statistics is necessary to make meaningful statements in the future. This may provide new insights into the temperature at which the onset of localization occurs.

\section{Influence of external magnetic fields with staggered quarks}\label{sec:influence_magnetic_fields}

\begin{figure}[!ht]
    \centering
    \begin{subfigure}{0.49\textwidth}
        \includegraphics[width=\linewidth]{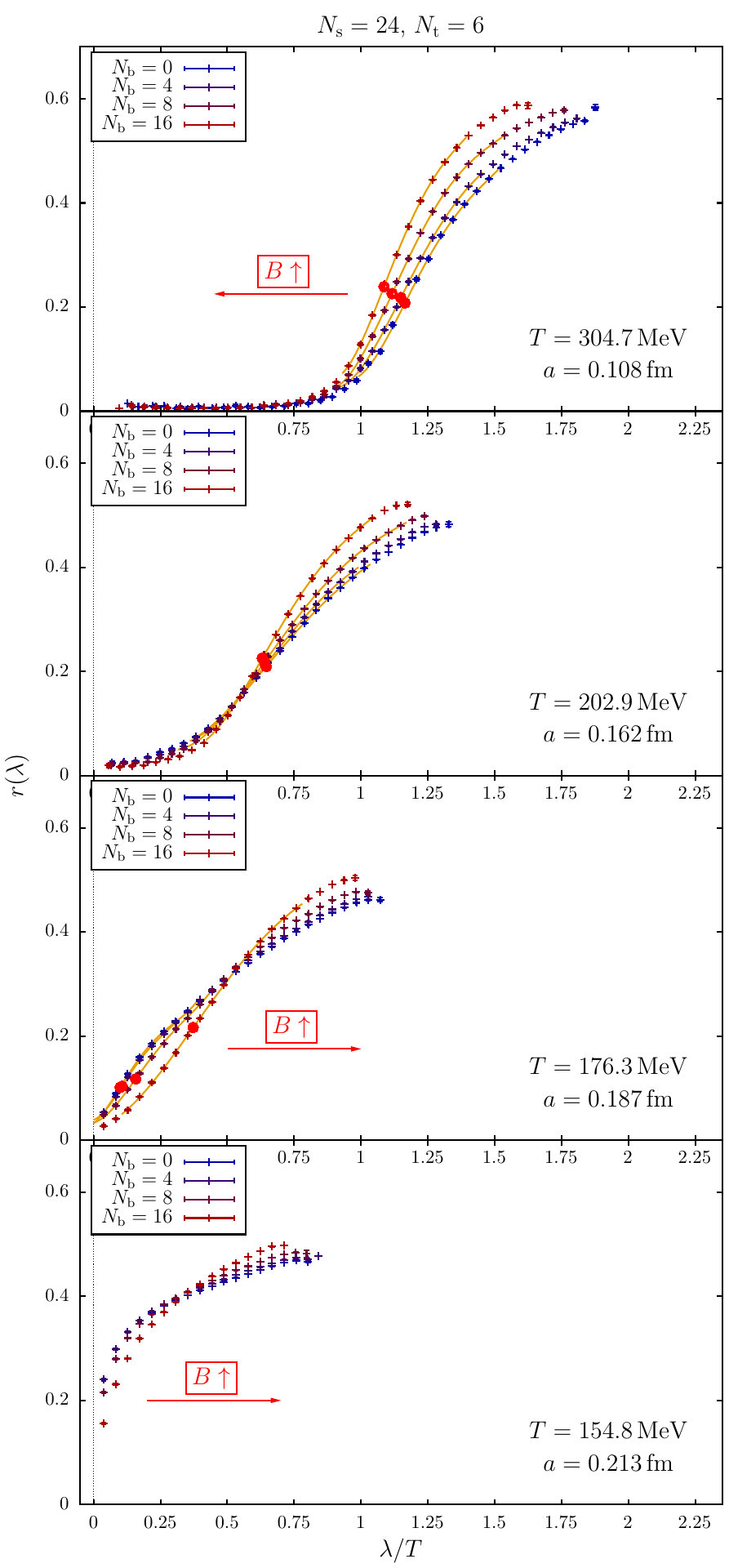} 
    \end{subfigure}
    \begin{subfigure}{0.49\textwidth}
        \includegraphics[width=\linewidth]{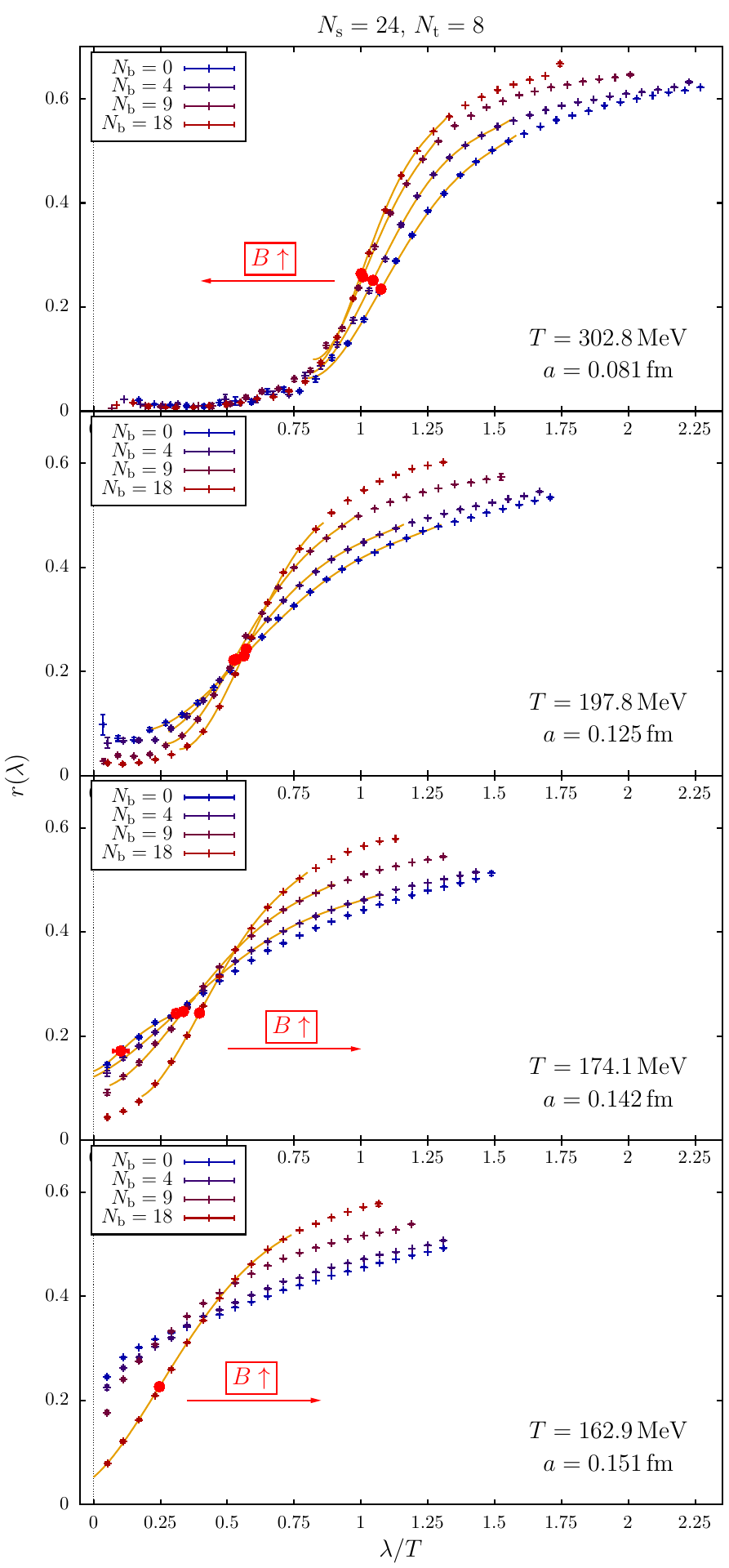} 
    \end{subfigure}
 
    \caption{Relative volume for $N_\mathrm{t}=6$ (left panels) and for $N_\mathrm{t}=8$ (right panels) as a function of the eigenvalues (shown as the dimensionless quantity $\lambda/T$) of the staggered Dirac operator for different magnetic field strengths and several temperatures. The mobility edge estimates are indicated by the red colored circles, which were determined by the displayed fitting curves. For the lowest temperature (bottom plots), the inflection point of $r(\lambda)$ disappears towards the left side. The small error bars at each point represent solely statistical errors. For visibility reasons, we only plot every third data point.}
    \label{fig:relvol_magnetic_field}
\end{figure}

In this section, we discuss our results on the influence of magnetic fields on the Anderson transition using a purely staggered setup. To this end, we use gauge configurations for two lattice spacings ($24^3\times 6$ and $24^3\times 8$) with $2+1$ flavors of staggered quarks with 2 steps of stout smearing in the fermion sector, the tree-level Symanzik-improved action in the gauge sector, and a uniform magnetic field background in the third spatial direction~\cite{Bali:2011qj}. In this setup, we compute the 400 lowest-lying eigenmodes of the staggered Dirac operator with a background magnetic field, which is given by
\begin{equation}
    D_\mathrm{stag}^f(n,l) = \sum_{\mu}\frac{\eta_{\mu}(n)}{2}\Bigl [U_{\mu}(n)\, u_{\mu}^f(n)\,\delta_{n+\hat{\mu},l}-U^{\dagger}_{\mu}(n-\hat{\mu})\,u_{\mu}^{f*}(n-\hat{\mu})\,\delta_{n-\hat{\mu},l} \Bigr] \,,  
    \label{eq:staggered_dirac_operator}
\end{equation}
where $\eta_{\mu}(n)$ are the staggered phases, and the flavor-dependent links $u^f_{\mu}(n)\in\mathrm{U(1)}$ implement the magnetic field. On a finite volume, the magnetic flux is quantized according to $eB = 6\pi N_\mathrm{b} / L^2$, where $e$ is the elementary charge, and $N_\mathrm{b}\in\mathbb{Z}$ is the magnetic flux quantum number. Unlike our setup in Sec~\ref{sec:overlap_tmft}, here, we vary the temperature via the inverse gauge coupling $\beta$. In Table~\ref{tab:overview_configurations}, we provide the simulation parameters for our evaluated ensembles (see also Ref.~\cite{Bali:2011qj} for further details). Analogously to Sec.~\ref{sec:overlap_tmft}, we compute the relative volume~\eqref{eq:eigenmode_relvol}, and determine the mobility edge using the procedure summarized in Sec.~\ref{sec:introduction}. In Fig.~\ref{fig:relvol_magnetic_field}, we show the relative volume for different temperatures and highlight the qualitative behavior of the mobility edge. To avoid substantial lattice artifacts associated with the magnetic field, we restrict the values of the magnetic flux quantum number to $N_\mathrm{b}\leq16$ for $N_\mathrm{t} = 6$, and $N_\mathrm{b}\leq18$ for $N_\mathrm{t} = 8$.

\begin{table}[!ht]
    \centering
    \begin{tabular}{|c|c|c|c|c|c|c|c|c|c|c|}
    \hline
    \multicolumn{11}{|c|}{$24^3 \times 6$} \\ \hline
        $\beta$ & 3.450 & 3.480 & 3.555 & 3.600 & 3.625 & 3.650 & 3.680 & 3.750 & 3.810 & 3.940 \\ \hline
        $T$ / $\mathrm{MeV}$ & 113.5 & 124.0 & 154.8 & 176.3 & 189.3 & 202.9 & 220.1 & 263.8 & 304.7 & 403.7 \\ \hline
        $a$ / $\mathrm{fm}$ & 0.290 & 0.265 & 0.213 & 0.187 & 0.174 & 0.162 & 0.149 & 0.125 & 0.108 & 0.081 \\
        \hline
        \multicolumn{11}{c}{}
    \end{tabular}
    \begin{tabular}{|c|c|c|c|c|c|c|c|}
    \hline
    \multicolumn{8}{|c|}{$24^3 \times 8$} \\ \hline
        $\beta$ & 3.650 & 3.675 & 3.700 & 3.725 & 3.750 & 3.775 & 
3.940 \\ \hline
        $T$ / MeV & 152.2 & 162.9 & 174.1 & 185.7 & 197.8 & 210.3 & 302.8 \\ \hline
        $a$ / fm & 0.162 & 0.151 & 0.142 & 0.133 & 0.125 & 0.117 & 0.081 \\ \hline
    \end{tabular}
    \caption{Overview of ensembles for both of our lattice spacings. We computed the eigenmodes of the staggered Dirac operator for several $N_\mathrm{b}$ in the range $[0,16]$ for $N_\mathrm{t}=6$, and in the range $[0,18]$ for $N_\mathrm{t}=8$. The number of evaluated configurations is of the order of $\approx 200-400$ for each choice of $\beta$ and $N_\mathrm{b}$.}
    \label{tab:overview_configurations}
\end{table}

Based on our results for $N_\mathrm{t}=6$, at a high temperature ($T=304.7\,\mathrm{MeV}$) the mobility edge decreases with $B$, as highlighted by the arrow in Fig.~\ref{fig:relvol_magnetic_field}. Interestingly, this behavior does not persist for lower temperatures. Instead, we observe that at $T=202.9\,\mathrm{MeV}$ -- around when the Banks-Casher gap closes -- the magnetic field has little influence on the mobility edge. Furthermore, at an even lower temperature ($T=176.3\,\mathrm{MeV}$), an opposite behavior sets in, and the mobility edge increases with $B$. We note that a further reduction of the temperature to $T=154.8\,\mathrm{MeV}$, which roughly coincides with the pseudocritical temperature at the physical point ($T_\mathrm{c} \approx 155$ MeV~\cite{HotQCD:2018pds,Borsanyi:2010bp}), results in the vanishing of all mobility edges. Nevertheless, the tendency of localization of the lowest-lying modes with the magnetic field still holds for lower temperatures.

The right-hand side of Fig.~\ref{fig:relvol_magnetic_field} shows comparable temperatures for the ensembles with $N_\mathrm{t}=8$, corresponding to smaller lattice spacings but larger volume. We observe a similar qualitative behavior of the mobility edge with $B$ at different temperatures. It is worth emphasizing that reducing the temperature to $T=162.9\,\mathrm{MeV}$, all mobility edges except the one for the strongest magnetic field vanish. This observation, combined with the behavior of the mobility edge at higher temperatures, suggests that the onset of localization, and thus the Anderson transition, occurs at a lower temperature in the presence of magnetic fields. This is analogous to what is known for the crossover temperature~\cite{Endrodi:2019zrl}. Therefore, it is tempting to hypothesize whether these phenomena could be related at a deeper level. Another way to clarify the possible relationship between the two is to study the pion mass dependence of the Anderson transition, particularly for heavy pions. In this regime, it is known that $T_\mathrm{c}$ is still reduced by $B$, whereas inverse magnetic catalysis disappears~\cite{Endrodi:2019zrl}. Nevertheless, we need finer lattices and, ultimately, a continuum extrapolation to make precise statements about the physics behind this behavior.

One possible speculation describing the behavior of the mobility edge could be the following: in a simplified picture where one thinks about the relative volume curves moving to smaller eigenvalues, this makes sense, since close to the chiral transition this movement has to slow down for magnetic fields, since otherwise the density of near-zero modes would increase producing a larger chiral condensate via the Banks-Casher relation \cite{Banks:1979yr}, which is however not allowed due to the inverse magnetic catalysis.

\section{Conclusion and outlook}\label{sec:conclusions}

In this proceedings article, we carried out two studies related to the Anderson transition using \gls{lqcd}, each of which we summarize in the subsections below. We also discuss prospects and research directions.

\subsection{\boldmath Overlap valence quarks on a twisted-mass sea at $B=0$}

In this first part, we provided updated results on the Anderson transition at $B=0$ using a mixed-action setup, where we computed the spectrum of the overlap operator on twisted-mass configurations. It is promising to further investigate the quantity $\tilde r$ as an alternative measure of localization, since it seems to be sensitive to the localization properties of the infrared modes, which are not captured by the relative volume (at least not without a volume scaling analysis). However, statistics need to be improved to enable an extraction of possibly two mobility edges via this quantity (c.f.\ Sec.~\ref{sec:overlap_tmft}). Therefore, an acceleration of the program computing the eigenmodes is desirable, which is ongoing and has shown some first promising tests. This also gives hope for the evaluation of the ensembles with a physical pion mass \cite{Kotov:2020hzm}, which is not feasible yet due to the immensely high computational costs coming with the increased number of lattice sites and density of low modes.

Furthermore, it is interesting to extend this study to other setups, namely, using other fermion discretizations. For example, we plan to compute overlap eigenmodes on domain-wall fermion configurations, such that the sea fermions have improved chiral properties, thus requiring smaller lattices due to the improved continuum limit.

\subsection{\boldmath Staggered setup at non-zero $B$}

In the second setup, we used purely staggered gauge configurations at non-zero $B$ both in the valence and in the sea sectors. We computed the relative volume and the mobility edge on two lattice sizes, namely, $24^3\times6$ and $24^3\times8$, for all our temperature and magnetic-field ensembles and found a non-monotonic behavior of the latter with the magnetic field on both ensembles. Specifically, at high $T$, the mobility edge decreases with $B$, whereas at low $T$, it decreases with $B$. At an intermediate temperature ($T \approx 200$ MeV), we found that the field has a negligible influence on the mobility edge. We speculate that this behavior change may be connected to the phenomenon of inverse magnetic catalysis, by which the chiral condensate is suppressed at temperatures near the crossover. Moreover, we found indications that the Anderson transition temperature might be reduced by the magnetic field. However, to make quantitative statements, a further extrapolation of the mobility edge to zero is necessary. 

Since we carried out our analysis using two relatively coarse lattice spacings ($N_\mathrm{t}=6$ and $N_\mathrm{t}=8$), further investigations are needed to shed light on the influence of external magnetic fields on the Anderson transition in continuum and infinite-volume QCD. For this reason, it is also relevant to quantify finite-volume effects on our observables. Furthermore, in a future publication, we aim at a more rigorous error analysis, including statistical and systematic contributions to the error budget of our observables. Therefore, we plan to apply Bayesian model averaging to capture these uncertainties more reliably, for instance, using the techniques developed in Refs.~\cite{Jay:2020jkz,Neil:2022joj}. Moreover, we comment that employing other measures of localization, like $\tilde r$ or \gls{ipr} ratios e.g.\ ${\rm IPR}_2(\lambda)/\sqrt{{\rm IPR}_3(\lambda)}$~\cite{Baranka:2023ani}, may be physically interesting as well.

\acknowledgments

The authors thank Gergely Endr\H{o}di for kindly providing the gauge configurations at non-zero magnetic field alongside the lattice code for the computation of the observables used in Sec.~\ref{sec:influence_magnetic_fields}, Bastian Brandt for granting computer time for this project at the Bielefeld cluster, and Matteo Giordano for discussions on the influence of magnetic fields on the Anderson transition. This work was partially supported by MICIU/AEI/10.13039/501100011033 and FEDER (EU) under Grant PID2022-140440NB-C21 and by Consejería de Universidad, Investigación y Innovación and Gobierno de España and EU -- NextGenerationEU, under Grant AST22 8.4. Robin Kehr and Lorenz von Smekal also acknowledge support by the Helmholtz Graduate School for Hadron and Ion Research (HGS-HIRe) for FAIR and the GSI Helmholtzzentrum f\"ur Schwerionenforschung.

\printglossary[type=\acronymtype]

\bibliographystyle{JHEP}
\bibliography{bibliography.bib}

\end{document}